\documentclass[twoside,journey]{IEEEtran}

\usepackage{makecell}
\usepackage{hyperref}
\usepackage{array}
\usepackage{graphicx,amssymb,amsmath}
\usepackage{multicol}
\usepackage[noadjust]{cite}
\usepackage{setspace}
\usepackage{subfigure}
\usepackage{graphicx}
\usepackage{float}
\usepackage {url}
\usepackage{stfloats}
\usepackage{amsthm,pifont}
\usepackage{cases,subeqnarray}
\usepackage{bm,multirow,bigstrut}
\usepackage{amsmath, amsthm, amssymb}
\usepackage{textcomp}
\usepackage{latexsym,bm}
\usepackage{booktabs}
\usepackage{xcolor}
\usepackage{mathtools}
\usepackage{dsfont}
\usepackage{extarrows}
\usepackage{epsfig}
\usepackage{flushend}
\usepackage{epstopdf}
\usepackage[noend]{algpseudocode}
\usepackage{algorithmicx,algorithm}
\theoremstyle{plain}

\theoremstyle{plain}

\usepackage{amsmath}

\IEEEoverridecommandlockouts
\begin{document}

\title{From Literature to Insights: Methodological Guidelines for Survey Writing in Communications Research}
\author{Dusit Niyato, \textit{Fellow, IEEE}, Octavia A. Dobre, \textit{Fellow, IEEE}, Trung Q. Duong, \textit{Fellow, IEEE},\\ George K. Karagiannidis, \textit{Fellow, IEEE}, and Robert Schober, \textit{Fellow, IEEE}

\thanks{D. Niyato is with the College of Computing and Data Science, Nanyang Technological University, Singapore (e-mail: dniyato@ntu.edu.sg). }
\thanks{O. A. Dobre is with the Department of Electrical and Computer Engineering, Memorial University, Canada (e-mail: odobre@mun.ca).}
\thanks{T. Q. Duong is with the Faculty of Engineering and Applied Science, Memorial University, Canada and also with the School of Electronics, Electrical Engineering and Computer Science, Queen's University Belfast, U.K. (e-mail: tduong@mun.ca).} 
\thanks{G. K. Karagiannidis is with the Department of Electrical and Computer Engineering, Aristotle University of Thessaloniki, Greece (e-mail: geokarag@auth.gr).}
\thanks{R. Schober is with the Institute for Digital Communications, Friedrich-Alexander University Erlangen-Nurnberg, Germany (e-mail: robert.schober@fau.de).}
}
\maketitle

\begin{abstract}
The rapid growth of communications and networking research has created an unprecedented demand for high-quality survey and tutorial papers that can synthesize vast bodies of literature into coherent understandings and actionable insights. However, writing impactful survey papers presents multifaceted challenges that demand substantial effort beyond traditional research article composition. This article provides a systematic, practical roadmap for prospective authors in the communications research community, drawing upon extensive editorial experience from premier venues such as the \textit{IEEE Communications Surveys \& Tutorials}. We present structured guidelines covering seven essential aspects: strategic topic selection with novelty and importance, systematic literature collection, effective structural organization, critical review writing, tutorial content development with emphasis on case studies, comprehensive illustration design that enhances comprehension, and identification of future directions. Our goal is to enable junior researchers to craft exceptional survey and tutorial articles that enhance understanding and accelerate innovation within the communications and networking research ecosystem.
\end{abstract}

\IEEEpeerreviewmaketitle

\section{Introduction}
Communications and networking technologies are evolving at tremendous speed \cite{7676258}. Innovations ranging from quantum communications and satellite-terrestrial integrated networks to neuromorphic computing architectures and early research into 7G are emerging faster than ever \cite{11016229, 11141666}. Accordingly, the number of research papers in these domains has increased exponentially over the past decade (see Fig.~1). For engineers, researchers, and students, keeping up with this torrent of new studies has become a formidable challenge. In this context, survey and tutorial papers have emerged as indispensable tools for navigating this deluge of information.

Unlike research articles that present original methodologies and results, surveys and tutorials synthesize and explain what is already established in a field. Specifically, a survey article offers a broad, comprehensive review of a specific topic, retracing its development from inception to the current state-of-the-art \cite{8368236}. In contrast, a tutorial article is designed to teach readers about a topic from the ground up, often adopting a more accessible, detailed, and step-by-step approach so that even non-experts can follow along \cite{10529221}. In other words, survey papers focus on \emph{what} has been accomplished in a domain, with liberal citation of the literature, while tutorial papers focus on \emph{how} to understand and apply concepts in that domain. Both types aim to educate readers by organizing complex topics into coherent narratives, albeit with different emphases and depths of technical detail. In recent years, the line between surveys and tutorials has begun to blur, giving rise to hybrid ``survey--tutorial" articles. This trend reflects a growing demand for papers that not only broadly cover the state-of-the-art but also explain it in depth, combining the strengths of both formats. Such hybrid articles provide immense value to the research community by simultaneously offering a high-level overview and a pedagogical introduction to a topic.
\begin{figure}[!t]
  \centering
  \includegraphics[width=1\linewidth]{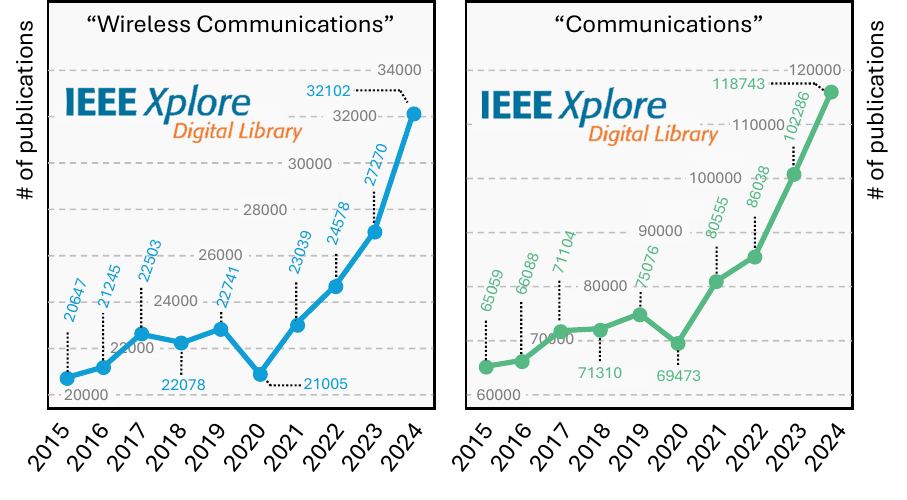}
  \vspace{-0.7cm}
  \caption{Publication trends in \textit{IEEE Xplore} database for communications research from 2015-2024. The left part shows the annual number of publications containing ``Wireless Communications'' in the title or keywords, while the right part displays publications related to ``Communications'' more broadly.}
  \vspace{-0.2cm}
\end{figure}

A well-crafted survey or tutorial paper enables readers to establish a comprehensive cognitive framework of the target field, effectively highlighting research gaps and suggesting directions for future work. Furthermore, it saves readers considerable time: instead of examining numerous separate papers to grasp a new topic, a survey systematically categorizes, reviews, and synthesizes related research into coherent themes. For newcomers, such papers are particularly valuable for gaining an in-depth understanding of a field's evolution and identifying key research trajectories. Consequently, the number of survey articles has been increasing in tandem with the explosion of primary research. High-quality surveys and tutorials have become essential components in the communications research ecosystem, as evidenced by their consistently high citation counts and widespread adoption as foundational references in subsequent studies. Therefore, there is a clear need for guidance on how to write effective survey and tutorial papers that stand out. 

Writing impactful survey or tutorial papers presents multifaceted challenges that extend even beyond research article composition. 
First, topic selection largely determines the paper's potential contribution and reception, requiring the authors of survey and tutorial papers to identify research domains that possess sufficient maturity for meaningful integration while avoiding areas already comprehensively covered by recent surveys. 
The scope definition process requires careful calibration: too narrow a focus may limit readership and impact, while overly broad coverage risks superficial description that fails to provide actionable insights. 
Once the boundaries are established, the authors face the critical task of comprehensive literature discovery across fragmented publication venues, particularly challenging in interdisciplinary areas where relevant contributions may span multiple research communities, such as IEEE. 
Moreover, the structural organization of surveys (usually 20-40 pages) is challenging, as the authors should determine logical frameworks for categorizing extensive literature, whether by methodology, chronology, application domains, or a combination thereof, while ensuring coherent progression through hundreds of papers. 
Throughout this process, maintaining narrative coherence while balancing technical depth with accessibility for a diverse readership requires sophisticated writing skills and deep domain expertise.
Last but not least, creating comprehensive visual representations through tables and figures that effectively synthesize quantitative findings and highlight key trends across the surveyed literature demands substantial effort and design expertise.

This article addresses these challenges directly by providing a structured, practical roadmap to prospective authors within the communications and networking community. Specifically, drawing upon the authors' extensive experience in survey/tutorial writing, editorial leadership of premier venues such as \textit{IEEE Communications Surveys \& Tutorials (COMST)}, \textit{IEEE Transactions on Network Science and Engineering (TNSE)}, \textit{IEEE Transactions on Communications (TCOM)}, \textit{IEEE Open Journal of the Communications Society (OJCOMS)}, and \textit{IEEE Communications Letters (COMML)}, and successful publication of highly-cited survey papers, this work systematically elucidates the fundamental aspects of effective survey and tutorial composition.

\begin{itemize}
   \item \textbf{Choosing the Right Topic and Scope}: Identifying timely research areas that offer significant interest and have yet to be comprehensively reviewed, and defining appropriate coverage of existing literature.
   \item \textbf{Organizing Content Effectively}: Strategies for structuring papers logically to guide readers seamlessly through background information, taxonomies, comparative analyses, and critical discussions.
   \item \textbf{Ensuring Depth and Insight}: Techniques for moving beyond mere literature summaries to include critical analysis, articulate open challenges, and synthesize significant trends and future directions.
   \item \textbf{Writing for Clarity and Impact}: Practical guidance on adopting an accessible writing style, clearly explaining complex terminology, and maintaining engagement to ensure comprehensibility even for readers outside the immediate research specialty.
\end{itemize}

We envision this article serving as a comprehensive resource for prospective survey and tutorial authors, providing the necessary guidance and support to craft exceptional contributions to the communications and networking literature. Our ultimate goal is to promote the development of high-quality surveys and tutorials that enhance understanding and accelerate innovation within our research community.


\section{Guidelines for Survey/Tutorial Writing}

\subsection{Topic Selection}
Selecting an appropriate topic is unarguably the most critical decision in survey writing, as it fundamentally determines the paper's potential impact and reception within the research community. The ideal survey topic should satisfy several key criteria that balance timeliness, importance, scope appropriateness, and differentiation from existing literature.

The most compelling survey topics address emerging technological trends or concepts that resonate with current research interests and practical needs. Recent examples include Beyond 6G communications, quantum computing for networking, large language models for communications, extremely large massive Multiple-Input Multiple-Output (MIMO), and near-field communications \cite{10685369, single2}. Moreover, topic selection must carefully consider the target journal's scope and readership expectations to ensure an appropriate fit. For \textit{IEEE COMST}, topics should focus primarily on communications systems, networking protocols, or closely related enabling technologies. Although interdisciplinary topics involving artificial intelligence, cybersecurity, and other domains can provide valuable insight, the communication and networking aspects should remain central to the discussion throughout the survey. Papers that focus primarily on pure technologies or general computer science topics without clear relevance to communication and networking may be considered out of scope.

Before committing to a specific topic, the authors should conduct a thorough analysis of existing survey literature. The proposed survey should offer clear added value through one or more of the following aspects: 1) broader scope coverage that encompasses recent developments, 2) deeper critical analysis that provides new insights, 3) a novel organizational perspective that improves accessibility, or 4) comprehensive treatment of emerging subfields not adequately addressed in existing surveys. The authors should prepare a detailed comparison table highlighting key differences from related surveys.

A practical consideration that significantly determines survey feasibility is the volume of available literature within the chosen scope. A comprehensive survey typically requires reviewing between 100 and 120 high-quality references to provide adequate coverage. Topics with insufficient literature (fewer than 50 significant papers) may indicate that the field remains too immature for comprehensive survey treatment, suggesting that the authors should either broaden the scope or wait for additional research development. Conversely, topics with overwhelming volumes of literature may require careful scope narrowing to maintain depth and analytical rigor. 

\subsection{Literature Collection}
The foundation of any high-quality survey lies in a systematic and comprehensive literature collection that ensures thorough coverage, credibility, and representativeness. 
For academic databases, \textit{IEEE Xplore Digital Library}\footnote{Available at: https://ieeexplore.ieee.org/Xplore/home.jsp} provides the most comprehensive coverage of communications and networking literature. 
Other publication databases can also offer strong computer science coverage that often includes relevant networking research. 
\textit{ArXiv}\footnote{Available at: https://arxiv.org/} contains numerous cutting-edge preprints that represent the latest developments in the field, although these papers have not undergone peer review and should be evaluated with appropriate caution.
\textit{Google Scholar}\footnote{Available at: https://scholar.google.com/} provides the broadest coverage, encompassing publications from major publishers, such as IEEE, along with \textit{ArXiv} preprints, doctoral dissertations, technical reports, etc. 
In addition, various scattered online resources such as industry datasets, standardization documents, technical white papers, and relevant news articles can serve as valuable supporting materials that provide practical context and real-world validation for academic research.

The quality of the collected papers should be carefully evaluated using multiple criteria to ensure the survey's credibility:
\begin{itemize}
    \item \textbf{Publication Venue}: Authoritative journals and conferences should constitute the backbone of survey references. High-impact venues include IEEE Transactions series, IEEE journals, IEEE magazine series, and premier international conferences. 
    \item \textbf{Citation Impact}: Papers with high citation counts (typically exceeding 100 citations) often demonstrate significant field impact, while ESI highly cited papers and IEEE popular articles represent works of exceptional influence that deserve careful consideration. 
    \item \textbf{Publication Timeliness}: Papers published within the last five years should form the core content of the survey, ensuring coverage of cutting-edge research directions. Simultaneously, earlier seminal works that established fundamental concepts, introduced breakthrough methodologies, or defined research paradigms should be selectively included to provide necessary theoretical foundations.
\end{itemize}

Manually managing hundreds of references is not recommended and can lead to potential oversights. Software packages such as Zotero\footnote{Available at: https://www.zotero.org/}, Mendeley\footnote{Available at: https://www.mendeley.com/}, or EndNote\footnote{Available at: https://endnote.com/} can facilitate efficient storage, categorization, and retrieval of references.

\subsection{Structure Organization}
Considering the typical length of survey papers, an effective structure is essential to maintain clarity and coherence. Most high-quality surveys begin with a comprehensive introduction that motivates the topic, defines the scope, situates the work relative to existing surveys, and explicitly highlights the key contributions. The second section is often of a preliminary or background nature. This part equips readers with the necessary foundations, such as basic theories, definitions, or terminology, before delving into more advanced discussions.
The following sections are usually the main body of the survey. Depending on the domain, this may take different forms:
\begin{itemize}
    \item \textbf{Function-centric Taxonomy:} The literature can be classified by functional tasks. For instance, the applications of AI in communication can be categorized into generation, classification, optimization, and prediction \cite{10685369}.
    \item \textbf{System-oriented Organization:} A layered or system-oriented organization can be adopted, where the literature is grouped by technical dimensions. For instance, the entire communication system can be divided into physical, networking, and application layers \cite{single2}.
    \item \textbf{Application-driven Structure:} In applications-rich domains, an application-driven structure is common, with separate sections devoted to verticals such as 5G, mobile-edge computing, and industry \cite{9861699}.
    \item \textbf{Framework-based Approach:} Finally, the authors can propose an explicit architectural framework at the beginning and analyze representative case studies through this lens. This is suitable for surveying emerging techniques with complicated components \cite{10242032}.
\end{itemize}
Beyond the core review, impactful surveys typically address cross-cutting issues that span multiple categories. Security and privacy are the most common, but datasets, benchmarks, industrial efforts, and standardization activities are also frequently discussed. Such transversal sections broaden the perspective and link academic research to practice.
Finally, a defining feature of high-quality surveys is the forward-looking component. Rather than scattering open questions throughout, leading surveys consolidate challenges and opportunities in a dedicated section. This overview not only identifies gaps but also outlines promising directions, providing readers with a roadmap for future research.

\subsection{Review Section Writing}
Writing effective review sections requires moving beyond simple literature summarization to provide genuine critical analysis that offers new insights and perspectives unavailable from reading individual papers in isolation.

Given that surveys often encompass hundreds of references, scattered individual paper reviews can easily overwhelm readers. Consequently, the authors should follow clear logical frameworks when organizing reviews. They can follow technical progressions, where works build upon each other sequentially to improve performance, or parallel approaches, where different papers tackle distinct tasks within a broader challenge. Moreover, the authors should employ various techniques to enhance analytical depth and clarity. For complex proposals and concepts, concrete examples help readers grasp abstract concepts and understand practical implementations (see the examples of Figs. 1 and 2 in \cite{8016573}). Additionally, comparative analysis between papers reveals relative strengths, weaknesses, and applicability conditions. 

After reviewing each group of literature, the authors should synthesize key insights that emerge from collective analysis rather than individual paper summaries. These insights might include common limitations shared across approaches, fundamental trade-offs that constrain all solutions, etc. For instance, reviewing blockchain consensus algorithms might reveal that performance improvements consistently require increased communication complexity \cite{10517413}. Such insights provide value beyond individual paper examination.

\subsection{Tutorial Section Writing}
The tutorial content demands fundamentally different writing approaches that prioritize pedagogy and progressive skills \cite{9968053}. The primary objective is to enable readers to understand and apply concepts effectively, regardless of their prior expertise in the specific domain.

The cornerstone of effective tutorial writing is step-by-step presentation that guides readers through complex concepts systematically. The authors should design clear learning progressions from basic concepts to complicated mechanisms and advanced applications, beginning each part with explicit learning objectives that tell readers what they should accomplish after completion. Complex topics should be broken into digestible subsections that build on each other, preventing cognitive overload while maintaining engagement. Throughout the presentation, various supporting elements enhance understanding: mathematical formulations should include detailed derivations with explanations for each transformation, schematic diagrams can illustrate abstract concepts visually, and algorithmic pseudocode can clarify procedural steps. 

Incorporating case studies in tutorial writing is strongly recommended due to two key advantages (e.g., in \cite{10529221}). First, concrete cases help readers better understand ``how to do'' rather than relying solely on abstract theoretical presentations, which can be difficult to comprehend without practical context. Second, case studies illustrate how tutorial methodologies solve real-world problems, enabling readers to envision applications within their own research scenarios. Authors should select cases that balance realism with accessibility, ensuring readers can follow the logical progression. 

\subsection{Illustration}
An often overlooked but crucial component of survey writing is the effective use of illustrations, including figures and tables. High-quality illustrations are not merely decorative. Instead, they serve as technical anchors that guide the reader through complex content and significantly improve comprehension. The frequently adopted illustrations in surveys can be categorized into five types.
\begin{itemize}
    \item \textbf{Survey Structure Figures}: These figures provide a visual roadmap of the entire survey, allowing readers to quickly capture organizational flow or adopted taxonomy (see Fig. 1 in \cite{10529221}).
    \item \textbf{Summary and Comparison Tables}: Such tables compress extensive literature into digestible formats, enabling side-by-side evaluation of approaches, methods, or systems from multiple perspectives (see Table V in \cite{10608156}). In particular, a comparison table in the Related Work part is highly effective for contrasting previous surveys with the current one, thereby highlighting novelty and contributions (see Table II in \cite{10242032}).
    \item \textbf{Technical Figures}: Diagrams such as system models, protocol stacks, or architectural frameworks serve as focal points in individual sections, facilitating a deeper understanding of complex mechanisms (see Figs. 9 and 12 in \cite{10242032}).
    \item \textbf{Experimental Results}: In domains with abundant experimental or empirical studies, performance curves, spectrum measurements, or latency comparisons provide clear evidence of trends, trade-offs, and limitations (see Figs. 6 and 7 in \cite{10529221}).
    \item \textbf{List of Abbreviations}: When a survey involves numerous abbreviations, a dedicated table improves readability and ensures quick checking (see \cite{9466363}).
\end{itemize}
Another important consideration is ensuring consistency between the illustrations and the main text. Figures and tables should not be taken in isolation: Their design and content must align closely with the narrative to avoid confusing readers. A common pitfall is to mention a figure or table briefly without integrating it into the discussion. 

\subsection{Identification of Challenges and Future Directions}
An indispensable component of a well-structured survey is the clear identification of open challenges and future research directions. This section is important not only for outlining the gaps in the existing body of work but also for ensuring that the survey provides constructive guidance to the research community. A crucial principle is that the discussion of future works should be closely aligned with the earlier structure of the paper: either reflecting the same taxonomy or offering complementary insights \cite{7081081}. Typically, 3-5 challenges and the corresponding future work can be considered from the following main perspectives and analyzed in this section.
\begin{itemize}
    \item \textbf{Technical Challenges}: Current solutions often face limitations in algorithmic performance, scalability, and efficiency. Future work can explore more effective algorithms, improved data availability, and energy-efficient designs to address these constraints.
    \item \textbf{Implementation and System Challenges}: Practical deployment is hindered by system complexity, hardware constraints, and interoperability issues. Future research can focus on lightweight architectures, real-time processing capabilities, and standardized interfaces to enhance scalability and integration.
    \item \textbf{Application Challenges}: Bridging the gap between theoretical advances and real-world deployment remains difficult. Future work can emphasize cross-domain validation, vertical-specific adaptations, and comprehensive benchmarking to accelerate adoption across industries.
    \item \textbf{Security and Privacy Challenges}: Increasing reliance on data-driven methods exposes systems to vulnerabilities such as adversarial attacks, data leakage, and lack of trust mechanisms. Future research should prioritize resilient security architectures, privacy-preserving computation, and trust management frameworks.
    \item \textbf{Standardization and Industrial Challenges}: The absence of unified protocols, benchmarks, and common platforms limits comparability and practical deployment. Future directions can focus on collaborative standardization, open-source platforms, and shared datasets to foster reproducibility and industry uptake.
\end{itemize}

\section{Discussion and Conclusion}
Writing survey and tutorial papers fundamentally differs from conducting original research, yet many junior researchers approach them with similar methodologies. The most common editorial observation involves treating surveys as simple literature compilations rather than scholarly synthesis that advances understanding. Truly impactful surveys do not merely catalog existing work; they reveal patterns, identify contradictions, and articulate insights that emerge only through comprehensive analysis. In this article, we provide an end-to-end guideline covering the principles and key considerations for each step of survey writing. We hope this systematic framework will enable the authors to transform their approach from passive literature compilation to active intellectual synthesis, ultimately producing survey and tutorial papers that contribute lasting foundational value to the communications and networking research community.

\section*{Acknowledgment}
We would like to sincerely thank Dr. Yinqiu Liu (e-mail: yinqiu001@e.ntu.edu.sg) from Nanyang Technological University, Singapore, for collecting and sorting the content of this paper.

\bibliographystyle{IEEEtran}
\bibliography{Ref}

\begin{thebibliography}{10}
\providecommand{\url}[1]{#1}
\csname url@samestyle\endcsname
\providecommand{\newblock}{\relax}
\providecommand{\bibinfo}[2]{#2}
\providecommand{\BIBentrySTDinterwordspacing}{\spaceskip=0pt\relax}
\providecommand{\BIBentryALTinterwordstretchfactor}{4}
\providecommand{\BIBentryALTinterwordspacing}{\spaceskip=\fontdimen2\font plus
\BIBentryALTinterwordstretchfactor\fontdimen3\font minus \fontdimen4\font\relax}
\providecommand{\BIBforeignlanguage}[2]{{%
\expandafter\ifx\csname l@#1\endcsname\relax
\typeout{** WARNING: IEEEtran.bst: No hyphenation pattern has been}%
\typeout{** loaded for the language `#1'. Using the pattern for}%
\typeout{** the default language instead.}%
\else
\language=\csname l@#1\endcsname
\fi
#2}}
\providecommand{\BIBdecl}{\relax}
\BIBdecl

\bibitem{7676258}
S.~M.~R. Islam, N.~Avazov, O.~A. Dobre, and K.-S. Kwak, ``Power-domain non-orthogonal multiple access ({NOMA}) in {5G} systems: Potentials and challenges,'' \emph{IEEE Communications Surveys \& Tutorials}, vol.~19, no.~2, pp. 721--742, Secondquarter 2017.

\bibitem{11016229}
U.~Khalid, U.~I. Paracha, Z.~Naveed, T.~Q. Duong, M.~Z. Win, and H.~Shin, ``Quantum fusion intelligence for integrated satellite-ground remote sensing,'' \emph{IEEE Wireless Communications}, vol.~32, no.~3, pp. 46--55, June 2025.

\bibitem{11141666}
P.~Jia, X.~Wang, Y.~Zhu, S.~Jin, and R.~Schober, ``Integrated heterogeneous service provisioning: Unifying beyond-communication capabilities with {MDMA} in {6G} and future wireless networks,'' \emph{IEEE Communications Magazine}, pp. 1--8, 2025.

\bibitem{8368236}
J.~Liu, Y.~Shi, Z.~M. Fadlullah, and N.~Kato, ``Space-air-ground integrated network: A survey,'' \emph{IEEE Communications Surveys \& Tutorials}, vol.~20, no.~4, pp. 2714--2741, Fourthquarter 2018.

\bibitem{10529221}
H.~Du, R.~Zhang, Y.~Liu, J.~Wang, Y.~Lin, Z.~Li, D.~Niyato, J.~Kang, Z.~Xiong, S.~Cui, B.~Ai, H.~Zhou, and D.~I. Kim, ``Enhancing deep reinforcement learning: A tutorial on generative diffusion models in network optimization,'' \emph{IEEE Communications Surveys \& Tutorials}, vol.~26, no.~4, pp. 2611--2646, Fourthquarter 2024.

\bibitem{10685369}
H.~Zhou, C.~Hu, Y.~Yuan, Y.~Cui, Y.~Jin, C.~Chen, H.~Wu, D.~Yuan, L.~Jiang, D.~Wu, X.~Liu, J.~Zhang, X.~Wang, and J.~Liu, ``Large language model ({LLM}) for telecommunications: A comprehensive survey on principles, key techniques, and opportunities,'' \emph{IEEE Communications Surveys \& Tutorials}, vol.~27, no.~3, pp. 1955--2005, June 2025.

\bibitem{single2}
D.~Fan, R.~Meng, X.~Xu, Y.~Liu, G.~Nan, C.~Feng, S.~Han, S.~Gao, B.~Xu, D.~Niyato, T.~Q.~S. Quek, and P.~Zhang, ``Generative diffusion models for wireless networks: Fundamental, architecture, and state-of-the-art,'' \emph{ArXiv preprint: ArXiv:2507.16733}, 2025.

\bibitem{9861699}
M.~M. Azari, S.~Solanki, S.~Chatzinotas, O.~Kodheli, H.~Sallouha, A.~Colpaert, J.~F. Mendoza~Montoya, S.~Pollin, A.~Haqiqatnejad, A.~Mostaani, E.~Lagunas, and B.~Ottersten, ``Evolution of non-terrestrial networks from {5G} to {6G}: A survey,'' \emph{IEEE Communications Surveys \& Tutorials}, vol.~24, no.~4, pp. 2633--2672, Fourthquarter 2022.

\bibitem{10242032}
C.~De~Alwis, P.~Porambage, K.~Dev, T.~R. Gadekallu, and M.~Liyanage, ``A survey on network slicing security: Attacks, challenges, solutions and research directions,'' \emph{IEEE Communications Surveys \& Tutorials}, vol.~26, no.~1, pp. 534--570, Fourthquarter 2024.

\bibitem{8016573}
Y.~Mao, C.~You, J.~Zhang, K.~Huang, and K.~B. Letaief, ``A survey on mobile edge computing: The communication perspective,'' \emph{IEEE Communications Surveys \& Tutorials}, vol.~19, no.~4, pp. 2322--2358, Forthquarter 2017.

\bibitem{10517413}
B.~Bellaj, A.~Ouaddah, E.~Bertin, N.~Crespi, and A.~Mezrioui, ``Drawing the boundaries between blockchain and blockchain-like systems: A comprehensive survey on distributed ledger technologies,'' \emph{Proceedings of the IEEE}, vol. 112, no.~3, pp. 247--299, March 2024.

\bibitem{9968053}
S.~Aboagye, A.~R. Ndjiongue, T.~M.~N. Ngatched, O.~A. Dobre, and H.~V. Poor, ``{RIS}-assisted visible light communication systems: A tutorial,'' \emph{IEEE Communications Surveys \& Tutorials}, vol.~25, no.~1, pp. 251--288, Firstquarter 2023.

\bibitem{10608156}
X.~Zhu, J.~Liu, L.~Lu, T.~Zhang, T.~Qiu, C.~Wang, and Y.~Liu, ``Enabling intelligent connectivity: A survey of secure {ISAC} in {6G} networks,'' \emph{IEEE Communications Surveys \& Tutorials}, vol.~27, no.~2, pp. 748--781, April 2025.

\bibitem{9466363}
Y.~Liu, K.~Qian, K.~Wang, and L.~He, ``Effective scaling of blockchain beyond consensus innovations and {M}oore's law: Challenges and opportunities,'' \emph{IEEE Systems Journal}, vol.~16, no.~1, pp. 1424--1435, March 2022.

\bibitem{7081081}
L.~Mohjazi, M.~Dianati, G.~K. Karagiannidis, S.~Muhaidat, and M.~Al-Qutayri, ``{RF}-powered cognitive radio networks: technical challenges and limitations,'' \emph{IEEE Communications Magazine}, vol.~53, no.~4, pp. 94--100, April 2015.

\end{thebibliography}

\end{document}